# Influence of La and Mn vacancies on the electronic and magnetic properties of LaMnO$_3$ thin films grown by pulsed laser deposition

Ivan Marozau[1], Proloy T. Das[1,2], Max Döbeli[3], James G. Storey[4], Miguel A. Uribe-Laverde[1], Saikat Das[1], Chennan Wang[1], Matthias Rössle[1], Christian Bernhard[1*]

[1] Department of Physics and Fribourg Center for Nanomaterials - Frimat, University of Fribourg, Chemin du Musee 3, CH-1700 Fribourg, Switzerland

[2] Department of Physics and Meteorology, Indian Institute of Technology, Kharagpur, PIN-721302, India

[3] Laboratory for Ion Beam Physics, ETH Zurich, Schafmattstrasse 20, CH-8093 Zurich, Switzerland

[4] Callaghan Innovation, P.O. Box 31310, Lower Hutt, New Zealand

[*]christian.bernhard@unifr.ch

## ABSTRACT

With pulsed laser deposition we have grown *c*-axis oriented thin films of the nominal composition LaMnO$_3$ (LMO) on LSAT(001) substrates. We find that, depending on the oxygen background pressure during growth, the LMO films contain sizeable amounts of La and/or Mn vacancies that strongly influence their electronic and magnetic properties. Specifically, we show that the Mn/La ratio can be systematically varied from 0.92 at 0.11 mbar to 1.09 at 0.30 mbar of oxygen. These cationic vacancies lead to markedly different disorder effects that become most pronounced once the samples are fully oxygenated and thus strongly hole doped. All as-grown and thus slightly oxygen deficient LMO films are ferromagnetic insulators with saturation moments in excess of 2.5 $\mu_B$ per Mn ion, their transport and optical properties can be understood in terms of trapped ferromagnetic polarons. Upon oxygen annealing, the most La-deficient films develop a metallic response with an even larger ferromagnetic saturation moment of 3.8 $\mu_B$ per Mn ion. In contrast, in the oxygenated Mn-deficient films the ferromagnetic order is almost completely suppressed to less than 0.5 $\mu_B$ per Mn ion and the transport remains insulator-like. We compare our results with the ones that were previously obtained on bulk samples and present an interpretation in terms of the much stronger disorder potential of the Mn vacancies as compared to the La vacancies. We also discuss the implications for the growth of LMO thin films with well-defined physical properties that, for example, are a prerequisite for the study of interface effects in multilayers.

# I. INTRODUCTION

The discovery of the so-called colossal magnetoresistance (CMR) effect in the perovskite-type manganate materials [1] has raised a strong interest in their versatile electronic and magnetic properties [2-6]. Most of these materials are derivatives of the rare earth (*RE*) manganates with the general formula *RE*MnO$_3$, that are doped on the *A*-site with lower valent cations, such as Ca$^{2+}$, Sr$^{2+}$ or Ba$^{2+}$ [3]. Depending on the *RE* element, as well as on the nature and the amount of the dopant the materials exhibit a great variety of structural, electrical and magnetic properties [4]. This leads to fascinating effects such as CMR [2,3,5], fully spin-polarized transport [4,7], a magnetic field induced insulator-to-metal transition [6], and multiferroicity [8]. It also gives rise to a variety of possible applications of the manganate perovskites as functional materials for magnetic sensors [9], magnetoresistive read-write heads for hard drives [2], random access memories [10,11], spintronics [12,13], catalysis [14-17], as well as in the field of fundamental materials research [18,19].

The most frequently used and well-explored family is lanthanum manganate doped on the *A*-site with alkali earth cations: La$_{1-x}$$A_x$MnO$_3$ (*A*: Ca, Sr, Ba). The desired properties for custom applications are usually obtained by varying the concentration of these divalent doping ions [4]. However, even for the parent compound with the nominal composition LaMnO$_3$ (LMO) it is well known that a strong variation of the electronic and magnetic properties can be achieved with defects such as Mn or La vacancies [20-24]. A prominent feature of LMO is its ability to accommodate a significant amount of cationic vacancies in addition to the oxygen vacancies which are common to many transition metal perovskites [25-31]. In truly stoichiometric LaMnO$_3$ all Mn ions have a single oxidation state of +3. The material is a Mott insulator and it exhibits a large Jahn-Teller distortion of the MnO$_6$ octahedra and a subsequent orbital order that leads to type-*A* antiferromagnetism with a Neel temperature of about 140 K [20,32]. In bulk LMO materials the cationic ratio of La/Mn ions is usually close to unity as it is predetermined by the stoichiometric mixture of the starting materials from which the samples are synthesized [20,21,26]. A significant concentration of La and Mn vacancies (in equal amounts) is still frequently found, especially in samples that are grown in an oxidizing atmosphere [20,21]. Often these are



expressed in terms of an excess oxygen content, δ, which represents the equal amounts of Mn and La vacancies and accounts for the resulting concentration of $Mn^{4+}$ ions [20, 24, 27, 29, 33]:

$$LaMnO_{3+\delta} = LaMn^{3+}_{(1-2\delta)}Mn^{4+}_{2\delta}O_{3+\delta} = \frac{3+\delta}{3}La_{3/(3+\delta)}Mn^{3+}_{(3-6\delta)/(3+\delta)}Mn^{4+}_{6\delta/(3+\delta)}O_3 \quad (Eq.\ 1)$$

In the following discussion, we do not use this notation in terms of the excess oxygen content, δ, but, for the sake of clarity, we rather quote explicitly the concentration of La and Mn vacancies. Often we will also refer to the corresponding concentration of $Mn^{4+}$ ions or the hole doping level, $p$, that is determined by the concentration of La and Mn vacancies. Depending on the preparation and annealing conditions, some LMO samples may also contain oxygen vacancies which counteract the hole doping effect of the cationic vacancies (they will mostly reside in unit cells with a missing La or Mn ion).

In the hole doped $La_{1-x}(Sr,Ca)_xMnO_3$ system the phase diagram of the electronic and magnetic properties is well established and evolves as follows. With increasing hole doping, $p$, the ferromagnetic moment increases steadily from about 0.2 $\mu_B$ per Mn ion at $p=0$, where it arises from a weak canting of the *A*-type antiferromagnetic order, to about 2-2.5 $\mu_B$ per Mn ion around $p \approx 0.1$-$0.15$ [4, 34, 35]. In this doping range the samples still exhibit a very high resistivity with an insulator-like temperature dependence that is commonly interpreted in terms of an activated polaronic transport mechanism [4]. This insulating ferromagnetic state is thought to arise from clusters around the $Mn^{+4}$ ions in which the Jahn-Teller distortion is replaced by a special kind of orbital order which enables a short-ranged double-exchange interaction [24, 36]. Upon further hole doping to $0.2 \leq p \leq 0.4$, as these ferromagnetic clusters develop a sizeable overlap, an itinerant long-range-ordered ferromagnetic state emerges in which the orbital order and the Jahn-Teller distortion are suppressed. The ferromagnetic moment reaches values close to the theoretical limit of 4-$p$ $\mu_B$ per Mn ion and the Curie temperature increases to values as high as $T^{Curie} \approx 270$ K and 370 K in $La_{0.67}Ca_{0.33}MnO_3$ and $La_{0.67}Sr_{0.33}MnO_3$, respectively [37]. This regime of the phase diagram is also well known for the CMR effect, which is a result of the competition between the long-range double-exchange interaction, favoring the itinerant ferromagnetic state, and the local Jahn-Teller distortions, tending to localize the charge carries thus



giving rise to an upturn of the resistivity with decreasing temperature in the paramagnetic regime at $T>T^{Curie}$.

Similar phase diagrams have been reported for LaMnO$_3$ samples in which the hole doping is caused by cationic vacancies according to the Eq. 1 [20, 24, 27, 29, 33]. However, in this case there exists an important distinction between the doping with La and Mn vacancies. The missing La$^{3+}$ or Mn$^{3+}$ ions both give rise to similar hole doping effects, but the Mn$^{3+}$ vacancies also act as strong localization centers for the holes and severely reduce the strength of the ferromagnetic exchange interaction [20, 21]. The latter effects can be understood in terms of the disruption of the network of Mn-O-Mn bonds which is essential for the charge transport and also for the ferromagnetic interaction based on the double-exchange mechanism.

In the insulating part of the doping phase diagram the disorder effect due to the Mn vacancies seems to be less important since they do not lead to fundamental changes of the electronic and magnetic properties. A sizeable ferromagnetic order develops here even in samples in which the hole doping is mainly caused by the Mn vacancies [21]. Albeit, the magnitude of the ferromagnetic moment is somewhat smaller than in samples that are predominantly doped with La vacancies [21]. On the contrary, at higher doping levels of $p>0.2$ the disorder effect of the Mn vacancies plays a very prominent role and gives rise to a fundamental change of the electronic and magnetic properties. In samples with a large amount of Mn vacancies the itinerant ferromagnetic phase is entirely absent and replaced with a glassy magnetic state which exhibits clear signatures of strong disorder and frustration [24]. The charge carriers in these samples remain localized [20, 21]. This is very different from the samples in which the same hole doping ($p>0.2$) is caused by a corresponding amount of La vacancies. Here the itinerant ferromagnetic state develops similar to the case of La$_{1-x}$(Ca,Sr)$_x$MnO$_3$ [21].

While this strong modification of the electronic and magnetic properties of the LMO parent compound is well documented for the case of bulk LMO samples [20-31, 33, 38], there exists a lack of systematic investigation on corresponding thin films. A number of publications report the fabrication of LaMnO$_{3+\delta}$ and La$_{1-y}$MnO$_{3\pm\delta}$ thin films by pulsed laser deposition (PLD) [39-44], molecular beam epitaxy [45], DC-magnetron sputtering [46], atomic layer deposition [47, 48], spin-coating [49], and spray-



pyrolysis techniques [50]. However, in most cases the cationic composition (i.e. the Mn/La ratio) has not been analyzed. Especially, for PLD grown films it is commonly assumed that the cationic composition matches that of the target [39-43]. While this is often the case for PLD grown thin films, there exist some important exceptions including the case of the LMO thin films [44].

In the following paper we show that the cationic composition of the PLD grown LMO films and their subsequent electronic and magnetic properties can be modified in a systematic way as a function of the deposition conditions. Specifically, we present a parametric study of the influence of the oxygen partial pressure during the PLD growth and we illustrate how the desired electronic and magnetic properties of the LMO films can be obtained by selecting proper growth conditions and post-deposition annealing treatment. Good agreement is obtained with the trends in bulk LMO samples suggesting that the vacancy engineering allows one to produce thin film with controlled properties. The latter may be used, for example, to study interaction effects and novel phenomena at the interfaces between different materials in multilayers and heterostructures [51-53]. The investigation of thin films can also help to avoid certain limiting factors and drawbacks that can occur in polycrystalline samples, e.g. due to grain boundary effects and segregation of secondary phases [21-23]. Finally, strain effects due to a lattice mismatch with the substrate can be used to tailor the properties of thin film materials [40, 54].

## II. EXPERIMENTAL

Thin films with a nominal composition of $LaMnO_{3+\delta}$ and thicknesses of about 100 nm were grown with the conventional pulsed laser deposition technique [55]. The samples were deposited using a KrF excimer laser ($\lambda$=248 nm, pulse duration of 25 ns) operating at a frequency of 2 Hz. A sintered ceramic pellet of stoichiometric $LaMnO_3$ with a diameter of 25 mm was used as a target. During the deposition the target was rotated and toggled with respect to the laser beam to ensure that each pulse interacted with a "fresh" surface on the target, and that no cumulative heating of the target by the laser pulses occurred. All films were deposited on (001)-oriented single crystalline $(LaAlO_3)_{0.3}(Sr_2AlTaO_6)_{0.7}$ (LSAT) substrates (5x5x0.5 mm$^3$). The parent compound ($LaMnO_3$) has a moderate negative lattice mismatch (~2%) to the substrate which enables an epitaxial growth of



slightly in-plane compressed films. Unlike the SrTiO$_3$ substrates, these LSAT substrates do not undergo any structural phase transitions at low temperatures that could impact the electronic and magnetic properties of the LMO films grown on top [56, 57].

At first we performed a preliminary parametric study on the influence of the deposition conditions, such as the substrate temperature, the laser fluence, and the substrate-to-target distance. The goal was to obtain thin films with a high structural quality that contain a sizeable amount of cationic vacancies. As an indication of the latter we used the ferromagnetic moment as deduced from the saturation value of the magnetization measured at 10 K. Based on the crystalline quality obtained from XRD, and the ferromagnetic moment of the LMO films we identified the following set of the best growth parameters that was used for the growth of all subsequent LMO films: laser fluence of 2.0 J cm$^{-2}$, substrate temperature of 825°C, target to substrate distance of 5.0 cm.

Using these conditions we have grown a series of LMO films for which the oxygen background gas pressure was varied between 0.11 and 0.30 mbar as detailed in Table 1. After the deposition the samples were rapidly cooled while leaving the oxygen partial pressure unchanged. After the determination of their structural, electronic and magnetic properties some of these as-grown LMO films were subject to a post-deposition annealing treatment. The latter was performed in a flowing oxygen atmosphere at 600°C for 24 hours to study the influence of the oxygenation. In a preliminary study we found that no noticeable uptake of oxygen occurs up to a temperature of 500°C whereas at 600°C the oxygen mobility in the LMO films is already fast enough such that a further increase of the annealing time or annealing temperature does not result in any further changes of the structural and the electromagnetic properties. This result is in line with previous findings, which suggest that the oxidation of slightly reduced LMO powders starts at about 550-600°C [20].

The film growth mode was monitored *in situ* by reflection high energy electron diffraction (RHEED) [58]. When possible (i.e. in the case of a predominant layer by layer growth mode) the sample thickness was evaluated from the temporal oscillations of the RHEED intensity [58].



The film topography was examined by atomic force microscopy (AFM) with an NT-MDT NTEGRA Aura microscope under ambient conditions. The root-mean-squared surface roughness ($R_q$) was calculated from AFM scans performed over an area of 5x5 µm² (512x512 points, scan speed of 1.5 µm s⁻¹).

The cationic composition, i.e. the Mn/La ratio, of the LMO films was obtained from Rutherford backscattering spectroscopy (RBS) [59]. This technique has a high sensitivity to the heavier elements, whereas an accurate determination of the oxygen content is difficult to achieve due to a lower sensitivity of the RBS technique to light elements. In addition, there is the problem of the overlap of the oxygen signals that arise from the oxide thin film and the oxide-based substrate. To avoid this overlap problem for the signals of the cations in the film (La and Mn) the samples for the RBS analysis were grown on MgO(001) substrates (using the same PLD conditions). The measurements were performed using a 2 MeV ⁴He ion beam and a silicon PIN diode detector that was positioned at an angle of 168°. The collected RBS data were simulated with the RUMP software [60]. The experimental uncertainty of the RBS analysis for the ratio of the cation concentration is about ±1-2%. However, we present the corresponding uncertainty interval for the Mn/La ratio of ±3% (Table 1 and Fig. 1), which accounts for the statistical variation of the cation stoichiometry between the samples from different batches deposited under the same conditions.

The film orientation, the crystallinity, and the phase purity were investigated with X-ray diffraction (XRD) analysis using a four-circle Rigaku SmartLab diffractometer equipped with a rotating anode source (9kW) yielding Cu-K$_\alpha$ radiation with a wavelength of λ=0.154059 nm, a parallel beam optics, and a double crystal Ge(220) monochromator. To determine the phase purity and the orientation of the films in the out-of-plane direction we performed θ-2θ scans in the range of 2θ=10° to 80° in steps of 0.01° (1° per minute). The film thickness ($d$) was calculated from the Kiessig oscillations around the (001) Braggs reflection. The accuracy for the thickness determination is about ±5%. The as-grown thin films have also been examined with in-plane φ-2θχ XRD scans in order to check their epitaxial growth. The data was acquired in the 2θχ range from 10° to 80° in steps of 0.008°



(1° per minute) at an incident angle θ of 0.4° with respect to the sample plane. The out-of-plane and in-plane unit cell parameters of the film were calculated using Braggs formula from the θ-2θ and φ-2θχ scans, respectively [61]. The experimental uncertainties are ±0.0003 nm for the out-of-plane and ±0.0005 nm for the in-plane lattice constants.

The transport properties of the LMO films were studied using a physical property measurements system (PPMS from Quantum Design) with an external Keithley 2602 System Source Meter. The resistance (*R*) was measured as a function of temperature (*T*) from 300 K down to 10 K at a cooling rate of 2 K/min using a four-probe DC technique. The data points were acquired in steps of 0.5 K. Each data point is the average of 100 fast single measurements. The measurements were performed in a constant current mode with a voltage limit of 100 mV. Depending on the sample resistance at room temperature, the driving current was varied from 100 nA to 10 μA. If the voltage limit was reached, due to a very high resistance, the measurement mode was automatically changed to a constant voltage mode in which the driving current was decreased until the minimum current was reached. The upper measurement limit for the resistance is about $10^9$ Ohm. The film resistivity (ρ) was calculated from the measured resistance and the film thickness, *d*, using the following equation for thin films, which assumes that the distance between the electrodes is much larger than *d*:

$$\rho = \frac{\pi}{\ln 2} \cdot R \cdot d \qquad (Eq.\ 2)$$

The magnetization was measured using the vibrating sample option of the PPMS system, with the external magnetic field (*H*) applied parallel to the sample plane. The magnetization was measured as a function of temperature as the film was cooled from 300 K to 10 K in a field of 1000 Oe. To determine the saturation value of the magnetization at 10 K magnetization loops were recorded for applied fields ranging from -1 to +1 Tesla. All magnetization data were corrected for the substrate contribution, which consists of a diamagnetic signal that is *T*-independent and an additional weak paramagnetic signal. The latter has been found to vary between different LSAT substrates but generally it has a noticeable effect on the signal only below ~30 K and at very high magnetic fields. The magnetic moments have been converted to units of Bohr magnetons ($\mu_B$) per Mn ion.

The room temperature thermoelectric power (*TEP*) was determined by measuring the voltage generated while a temperature gradient of approximately 2 K cm$^{-1}$ was maintained along the sample using a heat source of 100 mW at one end.

The optical properties of the LMO films have been investigated with a broad-band ellipsometry technique. The spectroscopic ellipsometry measurements were performed in the near-infrared to ultraviolet photon energy range of 0.5-6.5 eV with a commercial ellipsometer (WOOLLAM VASE) equipped with a UHV cryostat for a temperature range of 4 K<*T*<700 K. For the far-infrared (FIR) and mid-infrared (MIR) ranges we used a home-built setup as described in Ref. [62] that was either attached to a Bruker 66v spectrometer at the infrared beamline of the ANKA synchrotron at K.I.T Karlsruhe in Germany or to a Bruker 113v with a glowbar light source in our laboratory. To perform the substrate correction we model the ellipsometry data using the WOLLAM VASE software [63].

### III. RESULTS AND DISCUSSION

Pulsed laser deposition is a well-established thin film deposition technique that is especially suited for the growth of materials with a complex stoichiometry [55]. Due to the high energy of the laser photons (typical $h\nu$~5 eV) and the short pulse duration (~25 ns) the laser-target material interaction is predominantly photonic with only a minor degree of thermal heating [64]. Irrespective of the target´s chemical composition, this results in a congruent ablation (transfer of the target material into the plasma plume). This important advantage of the PLD technique enables one to achieve a stoichiometric deposition of thin films even for chemically complex materials [55]. Nevertheless, in some cases the stoichiometry of the film may still deviate from that of the target. One possible reason involves the complex dynamics of the ions in the plasma plume during the transfer from the target to the substrate. The most typical example concerns the loss of oxygen, which occurs since the light oxygen species are more strongly scattered on their way to the substrate than the much heavier cations [65]. This oxygen loss is usually compensated by performing the PLD growth in a background pressure of oxygen gas [55, 65], which yields an extra uptake of oxygen. An additional post-growth annealing treatment is often required to obtain the desired oxygen content in the film. Other reasons for off-stoichiometric samples can be the formation of strongly volatile species in the plasma plume [66] or the re-sputtering of certain elements, preferentially of the lighter elements, from the growing film that may be caused by the arriving highly energetic plasma species [65]. These processes depend on the conditions of the PLD growth, like the laser fluence and the size of the ablation area on the target, the distance between target and substrate, and the background gas pressure.

In the present study we have varied the oxygen gas pressure to tune the cation stoichiometry of the LMO films. All samples have been deposited by ablating from a nominally stoichiometric LaMnO$_3$ ceramic target with a Mn/La ratio close to unity. Only the background pressure of oxygen, $p(O_2)$, has been varied between 0.11 and 0.30 mbar (see Table 1). The RBS analysis confirms that the Mn/La ratio in these films exhibits a systematic variation as a function of the $p(O_2)$. As listed in Table 1 and shown in Fig. 1, the Mn/La ratio increases systematically with growing $p(O_2)$. At 0.11 mbar the samples are strongly Mn-deficient (Mn/La=0.92±0.03) while at 0.30 mbar they are substantially La-deficient (Mn/La=1.09±0.03). A Mn/La ratio close to unity is obtained for the samples grown at $p(O_2)$ of 0.20-0.25 mbar. We remark that with the RBS technique one can determine rather accurately the ratio of the Mn and La concentrations, whereas the determination of the oxygen stoichiometry is not possible with a sufficient accuracy due to the lower sensitivity of RBS to the light elements (like O) as well as due to the overlapping of the oxygen signal with the one of the substrate. Based on the RBS data alone we can therefore not tell whether in LMO films with a Mn/La ratio close to unity the Mn and La vacancies are absent or whether they are present in equal amounts. We can also not determine the amount of the oxygen vacancies in the as-grown films (although these are expected to be almost absent in the films that have been annealed in oxygen atmosphere).

To obtain a rough estimate of the concentration of the Mn and La vacancies in our LMO films, we compared their electronic and magnetic properties with those of the bulk LMO samples, for which the concentration of these cationic vacancies has been reported[20, 21, 24]. This comparison suggests that the hole doping level in the as-grown and the oxygen-annealed samples amounts to $p$~0.1 and $p$~0.25, respectively. From these hole doping levels, the determined Mn/La concentration ratio from RBS, and the assumption that the as-grown films contain a certain amount of oxygen vacancies that are removed during the $O_2$ annealing, we obtain the rough estimate of the composition of the LMO films that is summarized in Table 1.

These estimates establish the following trends: (i) all our PLD grown samples contain substantial amounts of cationic defects; (ii) the samples grown at the lowest $p(O_2)$ contain mostly Mn vacancies; (iii) for the samples grown at the highest $p(O_2)$ the vacancies are predominantly on the La sites; (iv) the as-grown samples contain a substantial amount of oxygen vacancies that get removed upon annealing in oxygen atmosphere at 600ºC.

The oxygen background pressure also has a strong influence on the growth mode and the microstructure of the films. The growth mode was monitored *in situ* during the PLD process by RHEED and the film microstructure was examined by AFM. The deposition of LMO at the lower $p(O_2)$ of 0.11 and 0.15 mbar results in a layer-by-layer two-dimensional growth and yields very smooth films with a low surface roughness on the sub-nanometer scale (see Table 1 and Figs. 2(a) and (b)). At higher oxygen background pressures of 0.20-0.25 mbar the growth mode changes and becomes spatially inhomogeneous with regions of two-dimensional and three-dimensional growth yielding a complex topography of the sample surface with an increased roughness (see Figs 2(c) and (d)). Finally, at 0.30 mbar the growth mode appears to be homogeneously three-dimensional. This results in a rough surface with a pronounced granular structure (see Fig. 2(e)). These changes of the growth mode as a function of $p(O_2)$ are indicative of a substantial variation in the mobility of the addions and the subsequent crystal lattice formation on the surface of the growing film. The two-dimensional growth at lower $p(O_2)$ suggests that the plasma species arriving at the surface have a sufficiently high energy and, thus, surface mobility to enable a layer-by-layer growth mode. These high energy ions may also give rise to a preferential re-sputtering of the lighter Mn atoms (but not of the heavier La) and thus may be the reason for the enhanced Mn deficiency of these films. The energy of the plasma species can be expected to decrease with increasing $p(O_2)$, which explains that the re-sputtering of Mn is reduced in these LMO films (Table 1 and Fig. 1). Likewise, the surface mobility of the ions is reduced such that the growth mode changes over from a two-dimensional to a three-dimensional one. The observed preferential La deficiency of the films deposited at 0.30 mbar (Table 1 and Fig. 1) remains a surprising result that cannot be explained by the re-sputtering scenario. A possible reason may be the formation of La-enriched volatile clusters in the plasma or at the surface of the growing film. Likewise, the LMO system may have an intrinsic tendency to form La vacancies rather than Mn vacancies.

The structural quality, the phase purity and the out-of-plane and in-plane lattice parameters of the LMO films have been characterized with XRD measurements. The out-of-plane θ-2θ diffraction patterns as shown in Fig. 3 reveal only the (00L) series of LMO reflections along with the corresponding reflections of the (001)-oriented LSAT substrate. This confirms that the films are fully $c$-axis oriented. The full-width at half maximum (*FWHM*) of the LMO (002) peaks is given in Table 1. It ranges from 0.09° to 0.19° ±0.03°, which is ~3-6 times larger than the *FWHM*=0.03° of the LSAT substrate (002) peaks. The difference is reasonably small indicating a good film crystallinity and homogeneous out-of-plane orientation. The smoother films grown at lower $p(O_2)$ reveal narrower Bragg peaks and therefore a better out-of-plane orientation as compared to the rougher films grown at higher pressures (Table 1). The presence of pronounced Kiessig fringes on the diffraction patterns of the LMO samples deposited at lower $p(O_2)$ (inserts in Fig. 3(a) and (b)) confirms their low roughness and uniform thickness. . The absence of any extra peaks in the diffraction patterns testifies for the phase purity of these films.

The in-plane ($a=b$) and the out-of-plane ($c$) lattice constants of the LMO films grown in different oxygen background pressures are shown in Table 1 and in Fig. 4(a). The in-plane cell parameters do not exhibit any pronounced dependence on $p(O_2)$ and do not change much after the post-deposition oxygen annealing. They range between 0.3895 and 0.3902 nm ±0.0005 nm, which is quite close to the in-plane lattice constant of the LSAT substrate of 0.3868 nm. This indicates a reasonably good lattice matching between the LSAT substrate and the deposited LMO films.

The out-of-plane cell parameters of the as-grown films are larger and range from 0.3924 to 0.3945 nm ±0.0003 nm (see Table 1). Fig. 4(b) shows that for all LMO films the post-growth oxygen annealing treatment at 600°C leads to a sizeable decrease of the $c$-axis parameter. Since the ionic radius of $Mn^{4+}$ is smaller than that of $Mn^{3+}$ [67], this is a clear indication that this annealing treatment gives rise to a significant uptake of oxygen and thus an enhanced hole doping effect. We assume that all O vacancies that are present in the as-grown samples are cured by this annealing treatment. This is supported by our finding that no further changes of the *c*-axis lattice constant or of the electronic and magnetic properties occur upon increasing the annealing duration or temperature. It is also noteworthy that the out-of-plane unit cell parameters for both as-grown and $O_2$-annealed films tend to increase with increasing the deposition pressure (i.e. with decreasing the amount of Mn vacancies) with the exception of the as-grown films at 0.30 mbar (see Fig. 4(b) and Table 1). This can be attributed to the disturbance of the of Mn-O-Mn network by the missing Mn species which leads to an overall decrease of the occupied crystal lattice volume and consequently to a more compact ion arrangement with decreased average bond distances. A corresponding decrease of the lattice constant with increasing the Mn-deficiency was previously also observed in bulk LMO samples [21].

Next we discuss the changes in electrical resistivity, $\rho$, of the LMO films as a function of the $p(O_2)$ during growth and the post-growth oxygen annealing treatment. The *T*-dependence of $\rho$ is displayed in Fig. 5(a) for the series of as-grown LMO films and in Fig. 5(b) for the same films after the oxygen annealing treatment. All as-grown samples in Fig. 5(a) exhibit an insulator-like behavior with a steep upturn of $\rho(T)$ below about 150 K towards very large values at low temperature that eventually exceed the upper sensitivity limit of our setup. The absolute values of $\rho(T)$ are rather similar for most of the films. The only exception is the strongly La-deficient film for which $\rho(T)$ is significantly lower and a shoulder-like feature develops below ~200 K. The latter is related to the ferromagnetic transition as is further discussed in the next paragraph in the context of the magnetization data. The oxygen annealing treatment leads to a significant decrease of the resistivity for all the LMO films. This is explicitly shown in Fig. 6, which directly compares the $\rho(T)$ curves for each LMO film in the as-grown and the oxygen-annealed states. For the Mn-deficient films in Figs. 6(a)-(c) the values of $\rho(T)$ decrease by about one order of magnitude, whereas the insulator-like shape remains almost unchanged. For the slightly La-deficient sample with a Mn/La ratio of 1.015 in Fig. 6(d), the decrease of $\rho(T)$ is already larger, but ρ(*T*) still exhibits an insulator-like upturn towards low-*T* (Fig. 6(d)). The most prominent changes occur for the strongly La-deficient film in Fig. 6(e) for which the resistivity in the oxygen-annealed state exhibits a pronounced decrease below about 200 K towards a strongly metallic state. The shape of the ρ(*T*) curve with a cusp around the Curie temperature of about 200 K (see the magnetization data in the next paragraph) and even the absolute low-*T* value of ρ(10 K)≈3*$10^{-3}$ Ω cm are similar to that of $La_{0.67}Ca_{0.33}MnO_3$ or $La_{0.67}Sr_{0.33}MnO_3$ [37], for which a combined metallic and ferromagnetic state develops below $T^{Curie}$ by the virtue of the long range double-exchange mechanism. These results demonstrate that the Mn and La vacancies, while giving rise to a similar hole doping levels, have widely different effects on the conduction mechanism in these LMO films. The doping on the La site, either by introducing La vacancies or by the substitution of La with Ca or Sr, apparently has very similar effects on the electronic properties and can be used to introduce a metallic state that is based on the so-called double exchange $Mn^{3+}$-O-$Mn^{4+}$ interactions. On the contrary, the doped holes introduced by the Mn vacancies remain strongly bound to these defects, which strongly disrupts the network of Mn-O-Mn bonds. As was already mentioned in the introduction, corresponding behavior in the transport properties has been observed for bulk LMO samples in Ref. [21] where it has also been interpreted in terms of a strong trapping of holes by Mn vacancies [20, 38].

The room temperature values of the thermoelectric power that was measured for some of the LMO films are summarized in Table 1. Their positive values confirm that the doped charge carriers have a hole-like character. The absolute values are larger for the Mn-deficient films than for the La-deficient ones; for all samples they are significantly lower after the oxygen annealing treatment which removes the oxygen vacancies. The *TEP* data thus confirms that the concentration and/or the mobility of the holes is significantly higher for the La-deficient films and increases upon oxygen annealing.

Next we discuss the magnetic properties of the LMO films. The thermal dependence of the field-cooled magnetization in an external field *H*=1000 Oe is shown in Fig. 7(a) for the as-grown LMO films and in Fig. 7(b) for the same samples after the oxygen annealing treatment. The comparison of the magnetization loops obtained at 10 K of the as-grown and the oxygen-annealed films is shown in

Fig. 8. All of the as-grown films exhibit a sizeable ferromagnetic component with a moment of about 2-2.5 $\mu_B$ per Mn ion. This confirms that they are located in the ferromagnetic insulator regime of the phase diagram and thus are substantially hole doped with $p\sim0.1$ [20, 21, 24] due to the cationic vacancies. The doping may be a bit higher for the more La-deficient samples as is suggested by the gradual increase of the Curie temperature, $T^{Curie}$, from about 120 K for the most Mn-deficient sample to about 180 K for the most La-deficient film. Although, a weak onset of the ferromagnetic signal around 200 K occurs even for the as-grown samples at $p(O_2) \leq 0.20$ mbar. Likewise, for the as-grown samples at 0.30 and 0.25 mbar the magnetization exhibits a weak decrease below about 100 K that is indicative of a small antiferromagnetic fraction. Both effects suggest that these films exhibit a certain degree of electronic inhomogeneity that likely arises from a variation in the local concentration of the Mn and La vacancies. In this context, one has to keep in mind that the Mn vacancies, in addition to their hole doping effect, have a destructive effect on the ferromagnetic order since they disrupt the network of Mn-O-Mn bonds through which the ferromagnetic double-exchange interaction proceeds. Also, for these as-grown samples the hole doping due to the cationic vacancies is still partially compensated by the counteracting effect of the oxygen vacancies.

The latter are removed in the oxygen-annealed LMO films. The decreasing values of the *c*-axis lattice constant, the electric resistivity, and the thermoelectric power all confirm that the oxygen annealing treatment leads to a significant increase of the oxygen content (it is expected to remove the oxygen vacancies that are created during the PLD growth but not to affect the cationic vacancies). Figure 7(b) shows that for the most La-deficient films this additional hole doping yields a significant increase in the magnitude of the ferromagnetic moment to about 3.8 $\mu_B$ per Mn ion. This value is close to the theoretical limit and to the maximal value that is observed in bulk $La_{0.67}(Ca,Sr)_{0.33}MnO_3$ [37]. In combination with the strongly metallic response below $T^{Curie}$ (as shown in Figs. 5(b) and 6(e)) this confirms that the most La-deficient samples are in the metallic ferromagnetic regime of the phase diagram at $p\sim0.25$. Accordingly, we can conclude that the concentration of cationic vacancies (mostly La vacancies) is about 8%.

As shown in Figs. 7(b) and 8(a)-(c), the corresponding oxygen-annealed LMO films which contain a large amount of Mn vacancies exhibit fundamentally different magnetic properties. As opposed to the further increase of the ferromagnetic moment in the strongly La-deficient sample, the ferromagnetic moment is now strongly suppressed from about 2.5 $\mu_B$ per Mn ion in the as-grown state to about 0.5 $\mu_B$ per Mn ion in the oxygenated state. Notably, this drastic suppression of the ferromagnetic order occurs despite the removal of the oxygen vacancies and the concomitant increase in hole doping. The evolution of this suppression of the static magnetic order as a function of the Mn/La ratio of the LMO films is summarized in Fig. 9, which compares the saturation magnetization at 10 K for the as-grown LMO films (solid circles) with the oxygen-annealed ones (open squares). A corresponding suppression of the ferromagnetic order in strongly hole doped Mn-deficient samples has been observed in bulk LMO [20, 21, 24]. A comparison with the phase diagram of Ref. [24] suggests that our Mn-deficient films have a hole doping level of $p\sim0.25$. The resulting estimate for the concentration of the cationic vacancies (these are now mostly Mn vacancies) is therefore similar to the strongly La-deficient samples and yields a lower limit of about 8%.

The drastic suppression of the ferromagnetic order in these Mn deficient samples upon an additional hole doping due to the removal of the oxygen vacancies remains to be understood. It may well reflect the different orbital states and competing interactions that are underlying the ferromagnetic orders in the insulating and the metallic states. In the insulating ferromagnetic state the doped holes are forming polaronic clusters with a short-ranged ferromagnetic exchange interaction around the $Mn^{4+}$ ions that is enabled by a special kind of local orbital order as described in Ref. [36]. The ferromagnetic order in these clusters seems to be very robust and only moderately affected by the Mn vacancies. Apparently, the destructive effect on the ferromagnetic order is strongly enhanced for the same amount of Mn vacancies as the hole doping is further increased. The main difference should be that the concentration of the ferromagnetic clusters increases and that they start to develop a sizeable overlap. Even if a further delocalization of the charge carriers and thus the formation of a truly itinerant state is inhibited by the Mn vacancies, it is surprising that the influence of the ferromagnetic double-exchange interaction is now drastically reduced. This suggests that the local orbital order around the $Mn^{4+}$ ions plays an important role in stabilizing the ferromagnetic order in the insulating part of the phase diagram. In the itinerant ferromagnetic state at higher doping it is known that the double-exchange interaction competes with the orbital order and especially with the Jahn-Teller distortion. It seems that the Mn vacancies weaken the long-ranged double-exchange interaction and tip the balance towards a localized state. However, the latter appears to have a different kind of short-range orbital order which does not support the ferromagnetic exchange interaction, not even on the local scale. Previous neutron studies and magnetization measurements on Mn-deficient bulk LMO samples have shown that this unusual magnetic state at high doping exhibits a glassy behavior that is characteristic of a strongly disordered and frustrated magnetic state [24].

We remark that even for the case of the strongly La-deficient films there exists a clear difference in the properties of the ferromagnetic state in the insulating regime at lower hole doping and the itinerant state at larger doping. This can be seen from Fig. 8(e) which compares the magnetization loops at 10 K for the sample in the as-grown and the oxygen-annealed states. In the low-field regime, the former exhibits a significantly wider hysteresis loop and a more gradual upturn of the magnetization towards high fields. This suggests a larger crystalline anisotropy with an easy axis that is not parallel to the external field (which is applied along the film plane). This characteristic difference is also consistent with the involvement of an orbital order in the insulating ferromagnetic state and its absence in the itinerant ferromagnetic state.

Finally, we discuss the optical data which confirm the significant differences in the electronic and magnetic properties of the La- and the Mn-deficient LMO films. Figure 10 shows an overview of the temperature dependence of the real part of the optical conductivity, $\sigma_1$, of the LMO films in the range of 0.01 to 6.5 eV. Previous optical studies on $La_{0.67}(Ca,Sr)_{0.33}MnO_3$ crystals have shown that the transition from a paramagnetic insulator or bad conductor to a ferromagnetic metal is accompanied by a sizeable redistribution of spectral weight from high energies above 2 eV to low energy where a Drude-like band develops [68, 69]. The spectral weight of this Drude-like response was found to roughly scale with the square of the ferromagnetic moment [68, 70]. This can be understood in terms of the competition between the local Jahn-Teller distortions and the double-exchange mechanism for which the strength of the ferromagnetic coupling is determined by the gain in the kinetic energy of the delocalized spin polarized charge carriers.

Figure 10 shows that such a characteristic redistribution of the spectral weight below $T^{Curie}$ from high to low energies also occurs in our ferromagnetic LMO films. The strongest effect is observed in the oxygen-annealed LMO film grown at 0.30 mbar in Fig. 10(b) for which a pronounced Drude-like peak develops below $T^{Curie} \approx 200$ K. An additional low-energy band, which appears around 0.3 eV, is likely due to polaronic effects. The spectral weight that is redistributed to below 0.8 eV into the Drude-peak and the polaronic band amounts to $\sim 4.6*10^6$ $\Omega^{-1}$ cm$^{-2}$ corresponding to a plasma frequency of $\omega_{pl} \approx 13200$ cm$^{-1}$. Assuming that the effective mass of the charge carriers equals the free electron mass, their effective concentration amounts to $N_{eff} \approx 1.88*10^{21}$ cm$^{-3}$ or a concentration of 0.12 electrons per Mn ion. Similar values have been derived for La$_{0.67}$(Ca,Sr)$_{0.33}$MnO$_3$ single crystals in the itinerant ferromagnetic state [68, 69, 71].

Figure 10(a) shows that a somewhat modified and weaker spectral weight transfer from high to low energy occurs for the 0.30 mbar LMO sample in the as-grown state. The conductivity above 2 eV is also noticeably reduced at $T<T^{Curie} \approx 180$ K and the missing spectral weight is redistributed to energies below 2 eV. However, all the spectral weight is accumulated in a finite-energy peak with a maximum around 0.65 eV and a second weaker maximum around 1.35 eV at 10 K. Both peaks are also present in the paramagnetic state at 200 and 300 K, albeit at a higher energy (by ~0.35 eV) and with a significantly reduced spectral weight. A corresponding behavior was previously observed for a La$_{0.9}$Sr$_{0.1}$MnO$_3$ single crystal from the insulating ferromagnetic part of the phase diagram [68]. The pronounced low energy band has been interpreted in terms of ferromagnetic polarons that are pinned to regions with a specific kind of local orbital order that favors the double-exchange interaction [24, 36]. Notably, the low energy tail of this relatively narrow band does not reach the origin and therefore does not contribute to the DC conductivity in the ferromagnetic state. This agrees with the insulator-like $T$-dependence of the dc resistance as shown in Fig. 6(e). The spectral weight of this polaronic band amounts to about $3*10^6$ $\Omega^{-1}$ cm$^{-2}$ yielding $N_{eff} \approx 1.23*10^{21}$ cm$^{-3}$ or 0.08 electrons per Mn ion. The ratio of the redistributed spectral weight of 3.0/4.6≈0.65 for the 0.30 mbar grown samples in the as-grown and the oxygen-annealed states roughly follows the one of the low-temperature ferromagnetic moment in Fig. 9 of $(3.7/4.1)^2$≈0.81. This is consistent with the point of view that in both samples, despite their different electronic properties, the double-exchange mechanism is at the heart of the ferromagnetic order.

Figure 10(c) shows that a qualitatively similar behavior is observed for the 0.11 mbar as-grown LMO film which is also in a ferromagnetic insulating state according to the resistance and magnetization data in Figs. 6(a) and 7(a). The lowest interband transition again exhibits a sizeable softening from 1.65 eV at 200 K to 1.4 eV at 10 K that is accompanied by a large spectral weight increase of about $1.7*10^6$ $\Omega^{-1}$ cm$^{-2}$ corresponding to $N_{eff}$≈$0.69*10^{21}$ cm$^{-3}$ or 0.045 electrons per Mn ion. This spectral weight is also accumulated from a broad range of frequencies between 2 and 6.5 eV. This is clearly seen in the conductivity difference spectra shown in Fig. 11(b). The ratio of the redistributed spectral weight to that of the 0.30 mbar O$_2$-annealed sample of 1.7/4.6≈0.37 also compares reasonably well with the ratio of the low-temperature ferromagnetic moments in Fig. 9 of $(2.9/4.1)^2$≈0.50.

Figure 10(d) shows the *T*-dependent spectral weight redistribution of the oxygen-annealed LMO film grown at 0.11 mbar for which the ferromagnetic order is strongly suppressed as shown in Figs. 7(b) and 8(a). Notably, even in this weakly ferromagnetic sample the optical spectra reveal a redistribution of spectral weight from high energies above 2 eV toward the lowest interband transition around 1.55 eV. The redistributed spectral weight amounts to ~$1*10^6$ $\Omega^{-1}$ cm$^{-2}$ yielding $N_{eff}$≈$0.4*10^{21}$ cm$^{-3}$ or ~0.025 electrons per Mn ion. The main difference with respect to the other three LMO films with sizeable ferromagnetic moments concerns the absence of a softening of this low-energy interband transition in the magnetic state. The mode gains spectral weight and eventually becomes a bit narrower with decreasing temperature but its center frequency remains almost unchanged.

We note that a similar behavior was previously observed in a detwinned LaMnO$_3$ single crystal [72] that was stoichiometric and thus close to the undoped state with a dominant *A*-type antiferromagnetic order. For light polarization along the *a-b* direction in which the spins are ferromagnetically coupled, it was found that the spectral weight of the lowest interband transition (now located around 2 eV) exhibits a sizeable spectral weight increase below $T^{\text{Neal}} \approx 140$ K. This spectral weight was shown to originate from a second band around 4 eV. The opposite trend with a spectral weight shift toward the high energy bands was observed for light polarization along the *c*-direction in which the spins couple in an antiferromagnetic fashion. This characteristic behavior has been interpreted in terms of a super-exchange model in which these bands have been assigned to intersite *d-d* transitions. The lowest band around 2 eV corresponds to the high spin transition between the $e_g$ orbitals that is enhanced by ferromagnetic spin correlations between the neighboring Mn sites. Three more bands in the range of 4 to 6 eV (shifted up by the Hunds-coupling) have been assigned to the corresponding low-spin transitions that are suppressed by the ferromagnetic spin correlations.

The similar trend observed in our $O_2$-annealed LMO film deposited at 0.11 mbar suggests that it may also be governed by *A*-type antiferromagnetic correlations. We note that our films are grown under slightly compressive strain. Accordingly, the antiferromagnetic order may be along the surface normal while the in-plane correlations are predominantly ferromagnetic. We emphasize that this LMO film is strongly hole doped and thus far from the undoped region at *p*<0.05 where the *A*-type antiferromagnetic order occurs in stoichiometric samples [20]. This hole doping is evident from the *TEP*, the decrease of the lattice parameter upon oxygen annealing and from the related decrease in resistance. It is also evident from the optical spectra as can be seen in Fig. 11(a), which compares the room temperature spectra of all four LMO films. The oxygen annealing gives rise to a pronounced red-shift of the low-energy bands in the range of 1 to 1.5 eV and 3 to 3.5 eV that correspond to the low- and high-spin intersite transitions between the $e_g$ levels. Such a red-shift of these bands upon increasing hole doping was also observed in $La_{1-x}(Ca,Sr)_xMnO_3$ single crystals where it was explained in terms of an overall decrease of the Madelung potential upon hole doping [68, 69, 71]. Even a similar decrease of the conductivity above 5 eV has been observed upon hole doping, which is thought to arise from a further broadening of a very broad peak with a maximum above 6.5 eV that is related to an on-site *p-d* transition between the oxygen and Mn bands.

The optical spectra also reveal important differences in these intersite *d-d* transitions between the most Mn- and the La-deficient films grown at 0.11 and 0.30 mbar of oxygen partial pressure, respectively. Figure 11(c) shows that for the as-grown samples the high-spin intersite transition between the $e_g$ levels is shifted from about 1.65 eV in the Mn-deficient samples (0.11 mbar) to about 0.95 eV in the La-deficient samples (0.30 mbar). In the latter, a remnant of the 1.65 eV peak is still visible, which suggests that in addition to the high concentration of La vacancies it contains a small amount of Mn vacancies. A similarly large shift is observed for the bands at 3.85 eV in the Mn-deficient samples and at 3.2 eV in the La-deficient samples that arise from the corresponding low-spin transitions. These sizeable energy differences are most likely caused by the very different effects that the Mn and La vacancies have on the spatially extended nature of the initial and final $e_g$ states. The considerable hardening for the case of the Mn vacancies confirms that they have a much stronger localization effect on the doped holes than the La vacancies.

## IV. CONCLUSIONS

In this paper we have illustrated that the cationic stoichiometry of $LaMnO_{3+\delta}$ films grown by PLD depends in a systematic way on the deposition conditions, namely on the oxygen partial pressure. Samples grown at $p(O_2) \leq 0.25$ mbar reveal a Mn-deficiency that increases with decreasing $p(O_2)$. The samples grown at $p(O_2)$ of 0.30 mbar are strongly La-deficient. The oxygen content in the films can be modified by the post-deposition annealing in flowing oxygen. The combination of these approaches allows one to prepare LMO samples with fully controlled chemical compositions that in turn determine the electrical and magnetic properties. It has been shown that depending on the cationic stoichiometry and oxygen content the studied LMO films exhibit similar trends in the properties reported for LMO powders: the Mn-deficient samples exhibit insulating thermal behavior of resistivity and weaker ferromagnetism than La-deficient samples, which have lower resistivity values and become metallic at low temperatures when sufficiently oxygenated. The obtained results represent an easy and versatile approach to tailor the properties of PLD-grown LMO thin films according to the requirements for specific applications.

**ACKNOWLEDGEMENTS**

We acknowledge Yves-Laurent Mathis for technical support during the measurements at the infrared beamline of the ANKA synchrotron at K.I.T. Karlsruhe in Germany. This work is supported by the Schweizer Nationalfonds (SNF) grant 200020-140225 and the National Centre of Competence in Research "Materials with Novel Electronic Properties - MaNEP". One of the authors (P.T.D.) would like to thank the Indo-Swiss Joint Research Program (ISJRP) and the Council of Scientific and Industrial Research (CSIR) for financial support.

# REFERENCES

1. R. von Helmolt, J. Wecker, B. Holzapfel, L. Schultz, and K. Samwer, Phys. Rev. Lett. **71**, 2331 (1993).
2. S. Jin, T. H. Tiefel, M. McCormack, R. A. Fastnacht, R. Ramesh, and L. H. Chen, Science **264**, 413 (1994).
3. A. P. Ramirez, J. Phys.-Condes. Matter **9**, 8171 (1997).
4. J. M. D. Coey, M. Viret, and S. von Molnar, Adv. Phys. **48**, 167 (1999).
5. A. J. Millis, Nature **392**, 147 (1998).
6. Y. Tokura and Y. Tomioka, J. Magn. Magn. Mater. **200**, 1 (1999).
7. J. H. Park, E. Vescovo, H. J. Kim, C. Kwon, R. Ramesh, and T. Venkatesan, Nature **392**, 794 (1998).
8. S. W. Cheong and M. Mostovoy, Nature Materials **6**, 13 (2007).
9. T. Venkatesan, M. Rajeswari, Z. W. Dong, S. B. Ogale, and R. Ramesh, Philosophical Transactions of the Royal Society a-Mathematical Physical and Engineering Sciences **356**, 1661 (1998).
10. Y. Fujisaki, Japanese Journal of Applied Physics **52**, 040001 (2013).
11. A. Asamitsu, Y. Tomioka, H. Kuwahara, and Y. Tokura, Nature **388**, 50 (1997).
12. W. R. Zhang, A. P. Chen, F. Khatkhatay, C. F. Tsai, Q. Su, L. Jiao, X. H. Zhang, and H. Y. Wang, Acs Applied Materials & Interfaces **5**, 3995 (2013).
13. M. P. de Jong, V. A. Dediu, C. Taliani, and W. R. Salaneck, J. Appl. Phys. **94**, 7292 (2003).
14. R. J. H. Voorhoeve, J. P. Remeika, and D. W. Johnson, Science **180**, 62 (1973).
15. R. J. H. Voorhoeve, J. P. Remeika, B. T. Matthias, and P. E. Freeland, Science **177**, 353 (1972).
16. E. G. Vrieland, J. Catal. **32**, 415 (1974).
17. J. H. Chen, M. Q. Shen, X. Q. Wang, G. S. Qi, J. Wang, and W. Li, Applied Catalysis B-Environmental **134**, 251 (2013).
18. J. Hoppler, et al., Nature Materials **8**, 315 (2009).
19. J. Chakhalian, et al., Nature Physics **2**, 244 (2006).
20. J. Topfer and J. B. Goodenough, J. Solid State Chem. **130**, 117 (1997).
21. J. Topfer and J. B. Goodenough, Chem. Mater. **9**, 1467 (1997).
22. P. A. Joy, C. R. Sankar, and S. K. Date, J. Phys.-Condes. Matter **14**, 4985 (2002).
23. P. A. Joy, C. R. Sankar, and S. K. Date, J. Phys.-Condes. Matter **14**, L663 (2002).
24. C. Ritter, M. R. Ibarra, J. M. DeTeresa, P. A. Algarabel, C. Marquina, J. Blasco, J. Garcia, S. Oseroff, and S. W. Cheong, Phys. Rev. B **56**, 8902 (1997).
25. B. C. Tofield and W. R. Scott, J. Solid State Chem. **10**, 183 (1974).
26. J. A. M. van Roosmalen, P. Vanvlaanderen, E. H. P. Cordfunke, W. L. Ijdo, and D. J. W. Ijdo, J. Solid State Chem. **114**, 516 (1995).
27. J. A. M. van Roosmalen, E. H. P. Cordfunke, R. B. Helmholdt, and H. W. Zandbergen, J. Solid State Chem. **110**, 100 (1994).
28. J. A. M. van Roosmalen and E. H. P. Cordfunke, J. Solid State Chem. **110**, 106 (1994).
29. J. A. M. van Roosmalen and E. H. P. Cordfunke, J. Solid State Chem. **110**, 109 (1994).
30. J. A. M. van Roosmalen and E. H. P. Cordfunke, J. Solid State Chem. **110**, 113 (1994).
31. J. A. M. van Roosmalen and E. H. P. Cordfunke, J. Solid State Chem. **93**, 212 (1991).
32. Y. Tokura and N. Nagaosa, Science **288**, 462 (2000).
33. J. H. Kuo, H. U. Anderson, and D. M. Sparlin, J. Solid State Chem. **83**, 52 (1989).
34. A. Urushibara, Y. Moritomo, T. Arima, A. Asamitsu, G. Kido, and Y. Tokura, Phys. Rev. B **51**, 14103 (1995).
35. H. Kawano, R. Kajimoto, M. Kubota, and H. Yoshizawa, Phys. Rev. B **53**, 2202 (1996).
36. R. Kilian and G. Khaliullin, Phys. Rev. B **60**, 13458 (1999).
37. G. J. Snyder, R. Hiskes, S. DiCarolis, M. R. Beasley, and T. H. Geballe, Phys. Rev. B **53**, 14434 (1996).
38. J. H. Kuo, H. U. Anderson, and D. M. Sparlin, J. Solid State Chem. **87**, 55 (1990).

39  A. Gupta, T. R. McGuire, P. R. Duncombe, M. Rupp, J. Z. Sun, W. J. Gallagher, and G. Xiao, Appl. Phys. Lett. **67**, 3494 (1995).
40  P. Murugavel, J. H. Lee, J. G. Yoon, T. W. Noh, J. S. Chung, M. Heu, and S. Yoon, Appl. Phys. Lett. **82**, 1908 (2003).
41  H. S. Kim and H. M. Christen, J. Phys.-Condes. Matter **22**, 146007 (2010).
42  W. S. Choi, et al., Journal of Physics D-Applied Physics **42**, 165401 (2009).
43  W. S. Choi, D. W. Jeong, S. Y. Jang, Z. Marton, S. S. A. Seo, H. N. Lee, and Y. S. Lee, J. Korean Phys. Soc. **58**, 569 (2011).
44  C. Aruta, et al., J. Appl. Phys. **100**, 023910 (2006).
45  P. Orgiani, C. Aruta, R. Ciancio, A. Galdi, and L. Maritato, Appl. Phys. Lett. **95**, 013510 (2009).
46  A. Kleine, Y. Luo, and K. Samwer, Europhys. Lett. **76**, 135 (2006).
47  K. Uusi-Esko and M. Karppinen, Chem. Mater. **23**, 1835 (2011).
48  H. Khanduri, et al., Journal of Physics D-Applied Physics **46**, 175003 (2013).
49  G. Kartopu and M. Es-Souni, J. Appl. Phys. **99**, 033501 (2006).
50  R. Todorovska, N. Petrova, D. Todorovsky, and S. Groudeva-Zotova, Appl. Surf. Sci. **252**, 3441 (2006).
51  M. Gibert, P. Zubko, R. Scherwitzl, J. Iniguez, and J. M. Triscone, Nature Materials **11**, 195 (2012).
52  T. Golod, A. Rydh, V. M. Krasnov, I. Marozau, M. A. Uribe-Laverde, D. K. Satapathy, T. Wagner, and C. Bernhard, Phys. Rev. B **87**, 134520 (2013).
53  D. K. Satapathy, et al., Phys. Rev. Lett. **108**, 197201 (2012).
54  J. H. Haeni, et al., Nature **430**, 758 (2004).
55  D. B. Chrisey and J. R. Huber, *Pulsed Laser Deposition of Thin Films* (John Wiley & Sons, Inc., New York, 1994).
56  J. Hoppler, J. Stahn, H. Bouyanfif, V. K. Malik, B. D. Patterson, P. R. Willmott, G. Cristiani, H. U. Habermeier, and C. Bernhard, Phys. Rev. B **78**, 134111 (2008).
57  V. K. Malik, et al., Phys. Rev. B **85**, 054514 (2012).
58  A. S. Arrot, in *Ultrathin Magnetic Structures I*, edited by J. A. C. Bland and B. Heinrich (Springer-Verlag, 2005), p. 177.
59  W. K. Chu, J. W. Mayer, and M. A. Nicolet, *Backscattering Spectrometry* (Academic Press, New York and London, 1978).
60  L. R. Doolittle, Nucl. Instrum. Methods Phys. Res., Sect. B **15**, 227 (1986).
61  W. H. Bragg and W. L. Bragg, Proc. R. Soc. London, A **88**, 428 (1913).
62  C. Bernhard, J. Humlicek, and B. Keimer, Thin Solid Films **455**, 143 (2004).
63  JA Woollam Co. Inc., http://www.jawoollam.com
64  P. R. Willmott and J. R. Huber, Rev. Mod. Phys. **72**, 315 (2000).
65  T. Lippert, in *Photon-based Nanoscience and Nanobiotechnology*, edited by J. J. Dubowski and S. Tanev (Springer, Amsterdam, 2006), p. 267
66  I. Marozau, A. Shkabko, M. Döbeli, T. Lippert, D. Logvinovich, M. Mallepell, C. Schneider, A. Weidenkaff, and A. Wokaun, Materials **2**, 1388 (2009).
67  R. D. Shannon, Acta Crystallogr., Sect. A: Found. Crystallogr. **32**, 751 (1976).
68  Y. Okimoto, T. Katsufuji, T. Ishikawa, T. Arima, and Y. Tokura, Phys. Rev. B **55**, 4206 (1997).
69  M. Quijada, et al., Phys. Rev. B **58**, 16093 (1998).
70  N. Furukawa, J. Phys. Soc. Jpn. **64**, 3164 (1995).
71  K. Takenaka, K. Iida, Y. Sawaki, S. Sugai, Y. Moritomo, and A. Nakamura, J. Phys. Soc. Jpn. **68**, 1828 (1999).
72  N. N. Kovaleva, A. V. Boris, C. Bernhard, A. Kulakov, A. Pimenov, A. M. Balbashov, G. Khaliullin, and B. Keimer, Phys. Rev. Lett. **93**, 147204 (2004).

# TABLE

Table 1. Deposition conditions, estimated composition and properties of the studied as-grown and oxygen-annealed nominal LaMnO$_3$ thin films: oxygen pressure during growth $p(O_2)$ and post-growth annealing treatment (as-grown or annealed in oxygen); Mn/La ratio determined by RBS; estimated film composition*; root-mean-squared roughness $R_q$; in-plane cell parameter $b$; out-of-plane cell parameter $c$; full-width at half maximum (*FWHM*) of the (002) LMO film Bragg-peak; thermoelectric power *TEP*.

| $p(O_2)$, mbar; annealing | Mn/La ratio ±3% | estimated composition* | $R_q$, nm ±20% | $b$, nm ±0.0005 nm | $c$, nm ±0.0003 nm | FWHM ±0.03,° (002) peak | TEP, µV K$^{-1}$ ±1% |
|---|---|---|---|---|---|---|---|
| 0.11, as-grown | 0.92 | La$_{1.00}$Mn$_{0.92}$O$_{2.93}$ | 0.5 | 0.3898 | 0.3924 | 0.09 | +117 |
| O$_2$-ann. | | La$_{1.00}$Mn$_{0.92}$O$_{3.00}$ | | | 0.3862 | | +26.6 |
| 0.15, as-grown | 0.94 | La$_{0.99}$Mn$_{0.93}$O$_{2.93}$ | 0.7 | 0.3895 | 0.3925 | 0.09 | |
| O$_2$-ann. | | La$_{0.99}$Mn$_{0.93}$O$_{3.00}$ | | | 0.3866 | | |
| 0.20, as-grown | 0.98 | La$_{0.97}$Mn$_{0.95}$O$_{2.93}$ | 1.7 | 0.3900 | 0.3943 | 0.17 | |
| O$_2$-ann. | | La$_{0.97}$Mn$_{0.95}$O$_{3.00}$ | | | 0.3876 | | |
| 0.25, as-grown | 1.02 | La$_{0.95}$Mn$_{0.97}$O$_{2.93}$ | 2.2 | 0.3901 | 0.3945 | 0.19 | |
| O$_2$-ann. | | La$_{0.95}$Mn$_{0.97}$O$_{3.00}$ | | | 0.3876 | | |
| 0.30, as-grown | 1.09 | La$_{0.92}$Mn$_{1.00}$O$_{2.93}$ | 6.0 | 0.3902 | 0.3925 | 0.14 | +70.0 |
| O$_2$-ann. | | La$_{0.92}$Mn$_{1.00}$O$_{3.00}$ | | | 0.3884 | | +8.13 |

*The composition is roughly estimated based on the following:

- the Mn/La concentration ratio as determined by RBS;

- the hole doping levels in the as grown and oxygen-annealed samples of 0.1 and 0.24, respectively, as deduced from the comparison of their electronic and magnetic properties with those of bulk LMO samples;

- the assumption that the as grown samples contain similar amounts of the oxygen vacancies that are removed during the oxygen annealing treatment.

**FIGURES**

Figure 1. Variation of the Mn/La ratio in the nominal $LaMnO_3$ films as obtained from RBS as a function of the oxygen background pressure during deposition.

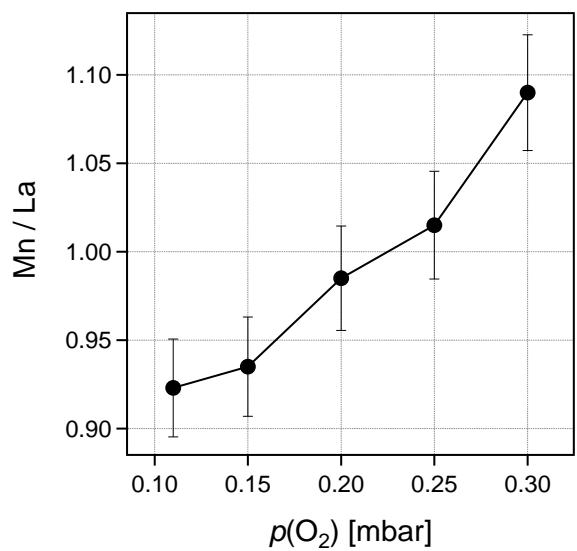

Figure 2 (Color online). AFM images of the as-grown LMO films in: **(a)** 0.11 mbar $O_2$; **(b)** 0.15 mbar $O_2$; **(c)** 0.20 mbar $O_2$; **(d)** 0.25 mbar $O_2$; **(e)** 0.30 mbar $O_2$. The insets give the root-mean-squared roughness.

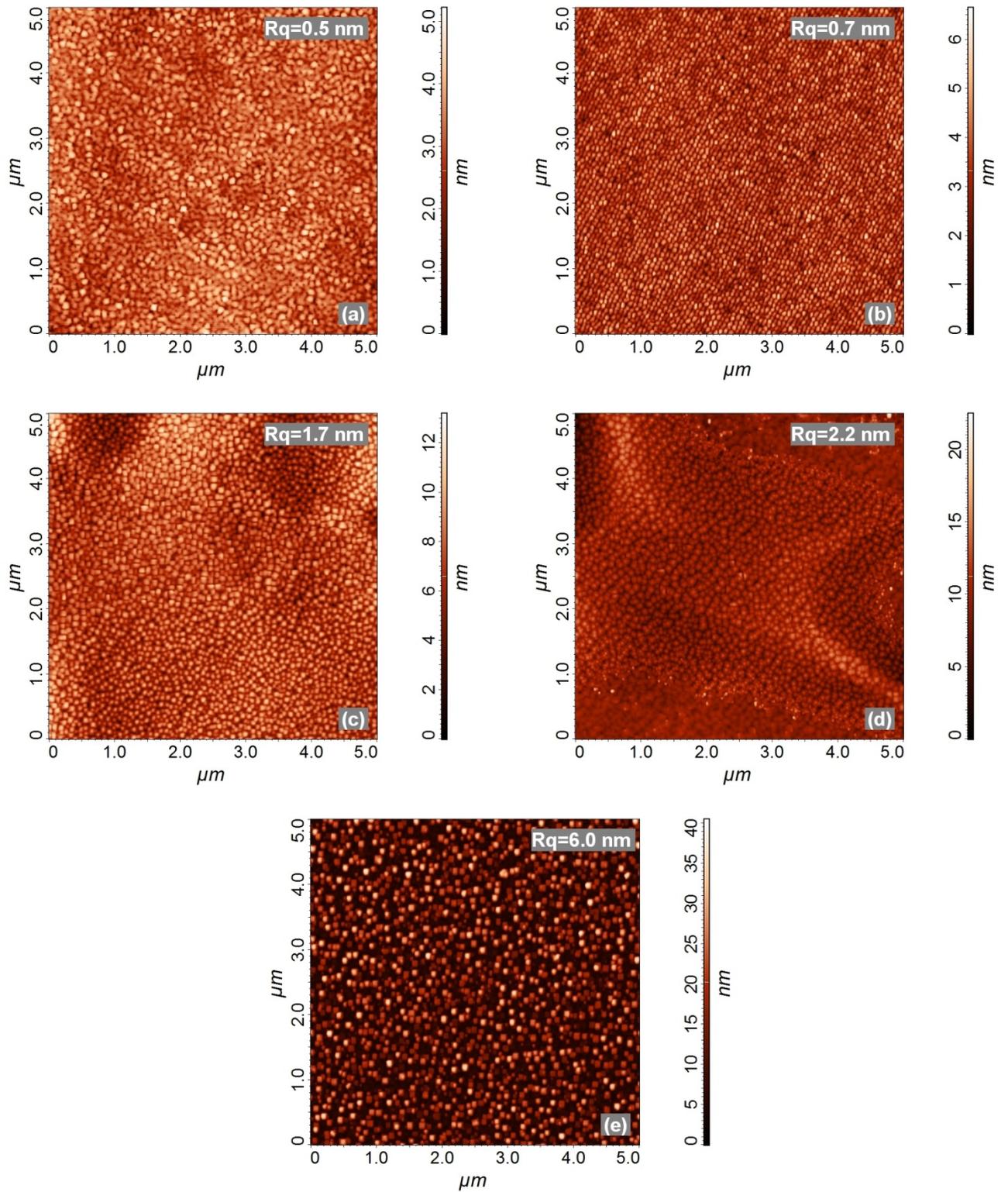

Figure 3 (Color online). Out-of-plane XRD patterns of the as-grown LMO films. The presence of only the (00*L*) Bragg-peaks of LMO confirms the *c*-axis oriented growth on the LSAT(001) substrates. The inserts detail the region around the (001) Bragg peak of LMO where the appearance of Kiessig fringes indicates a low roughness and homogeneous film thickness. Their periodicity has been used to determine the film thickness.

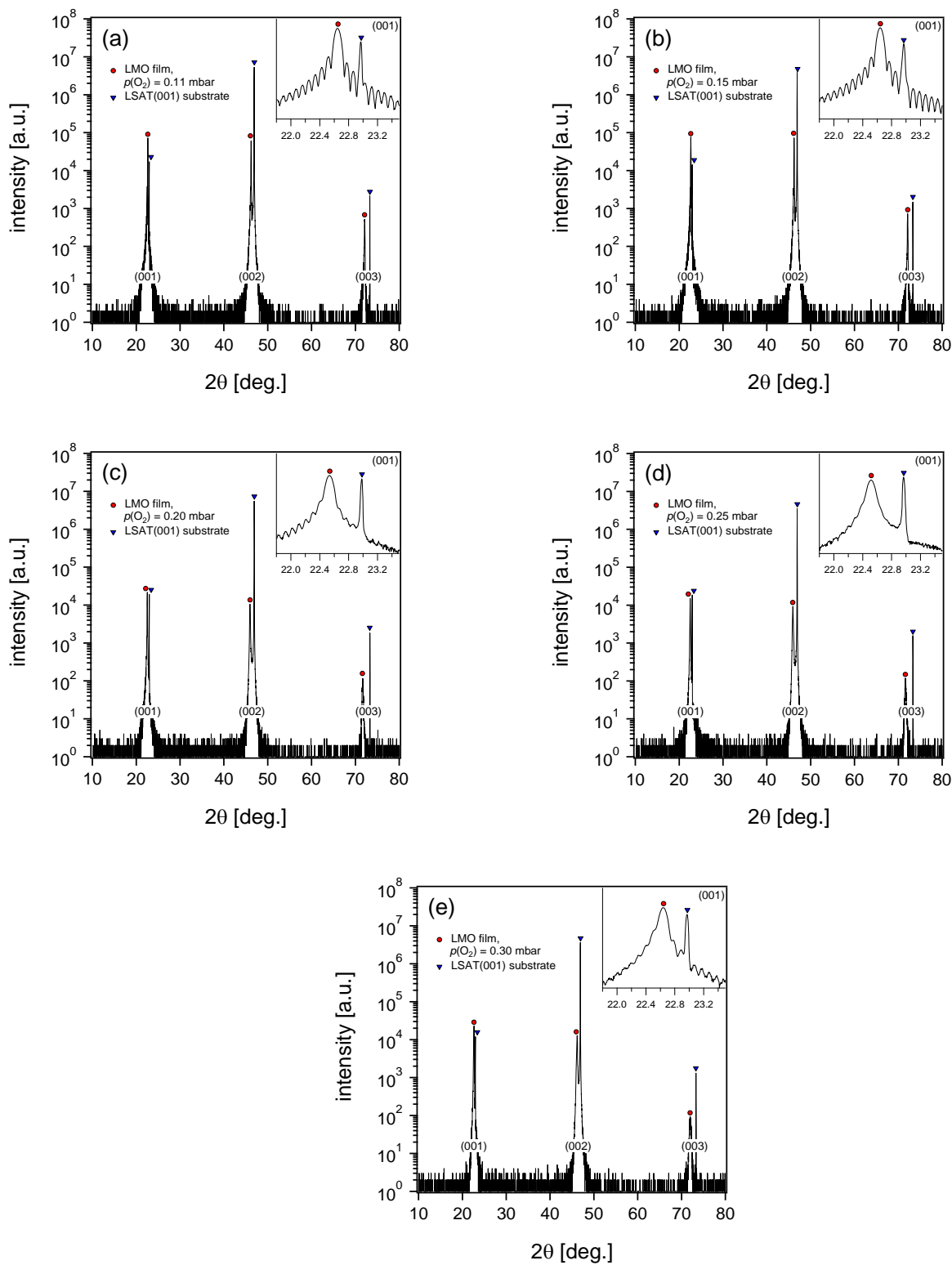

Figure 4 (Color online). **(a)** Variation of the in-plane and out-of-plane lattice constants of the as-grown LMO films as a function of the oxygen background pressure during deposition. **(b)** Comparison of the out-of-plane lattice parameters before and after the post-growth oxygen annealing treatment of the films

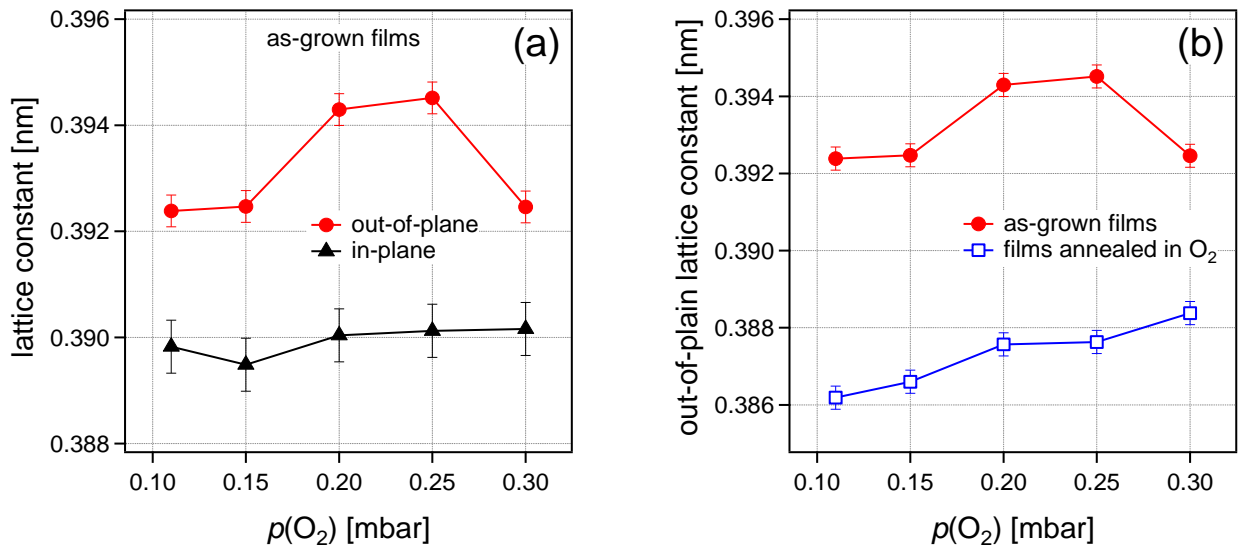

Figure 5 (Color online). Temperature dependence of the resistivity of the LMO films: **(a)** as-grown samples; **(b)** oxygen-annealed samples. All samples except for the most La-deficient film after the oxygen annealing treatment exhibit an insulator-like behavior. The latter becomes metallic below 200 K.

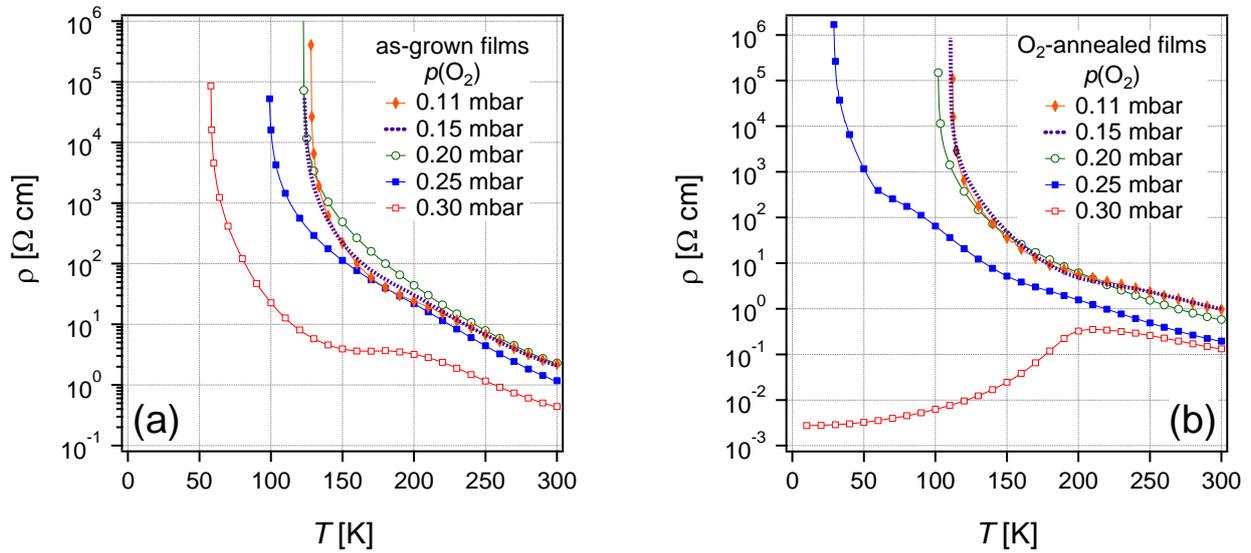

Figure 6 (Color online). Temperature dependence of the resistivity of the LMO films before and after the post-growth oxygen annealing treatment.

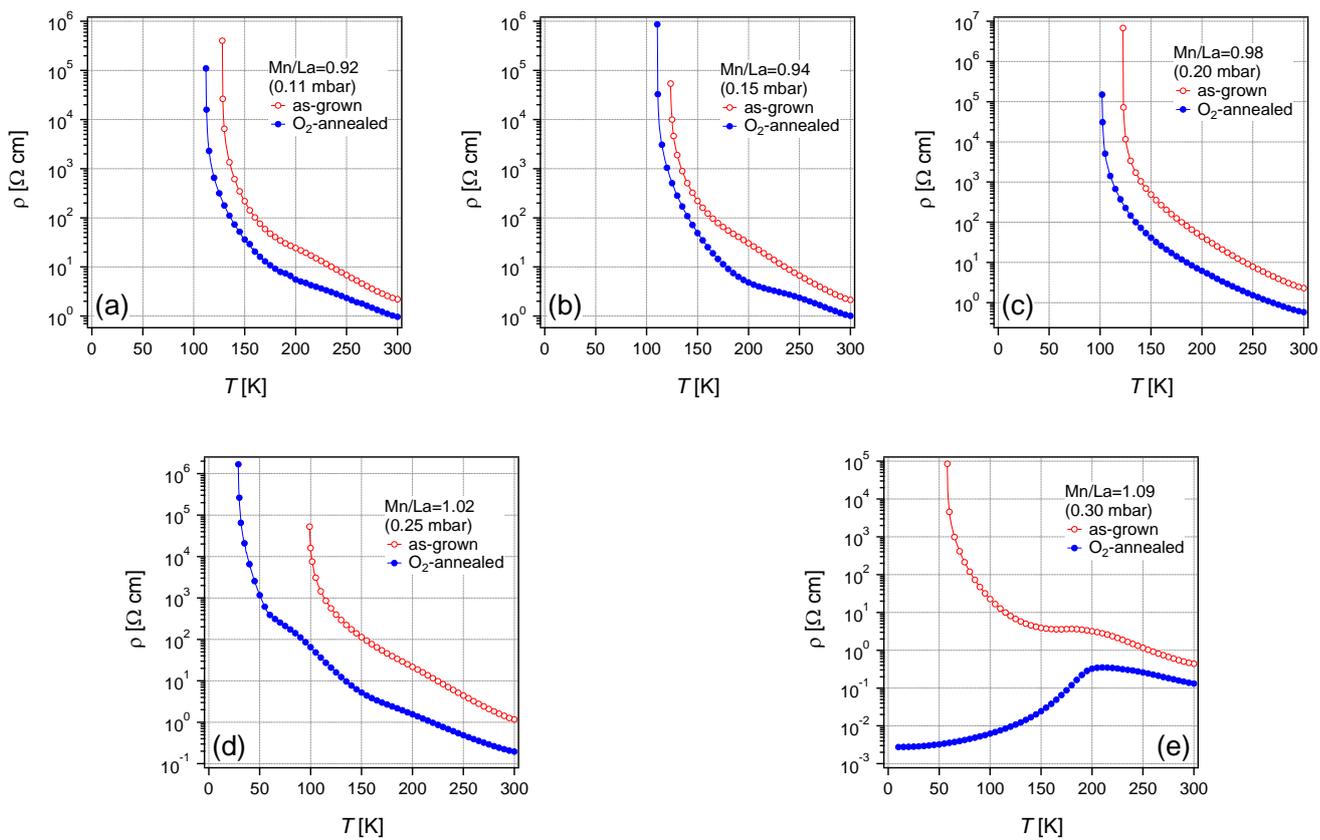

Figure 7 (Color online). Temperature dependence of the magnetization of the LMO films measured in field-cooled mode with an external field of 1000 Oe: **(a)** as-grown samples; **(b)** oxygen-annealed samples.

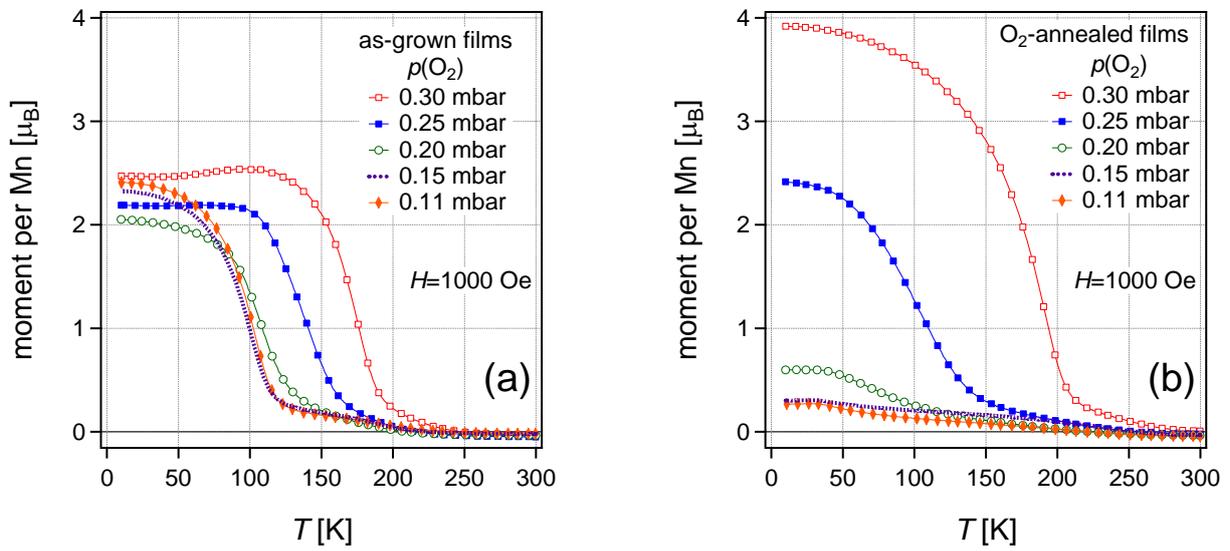

Figure 8 (Color online). Magnetization loops of the LMO films measured at $T$=10 K before and after the oxygen annealing treatment.

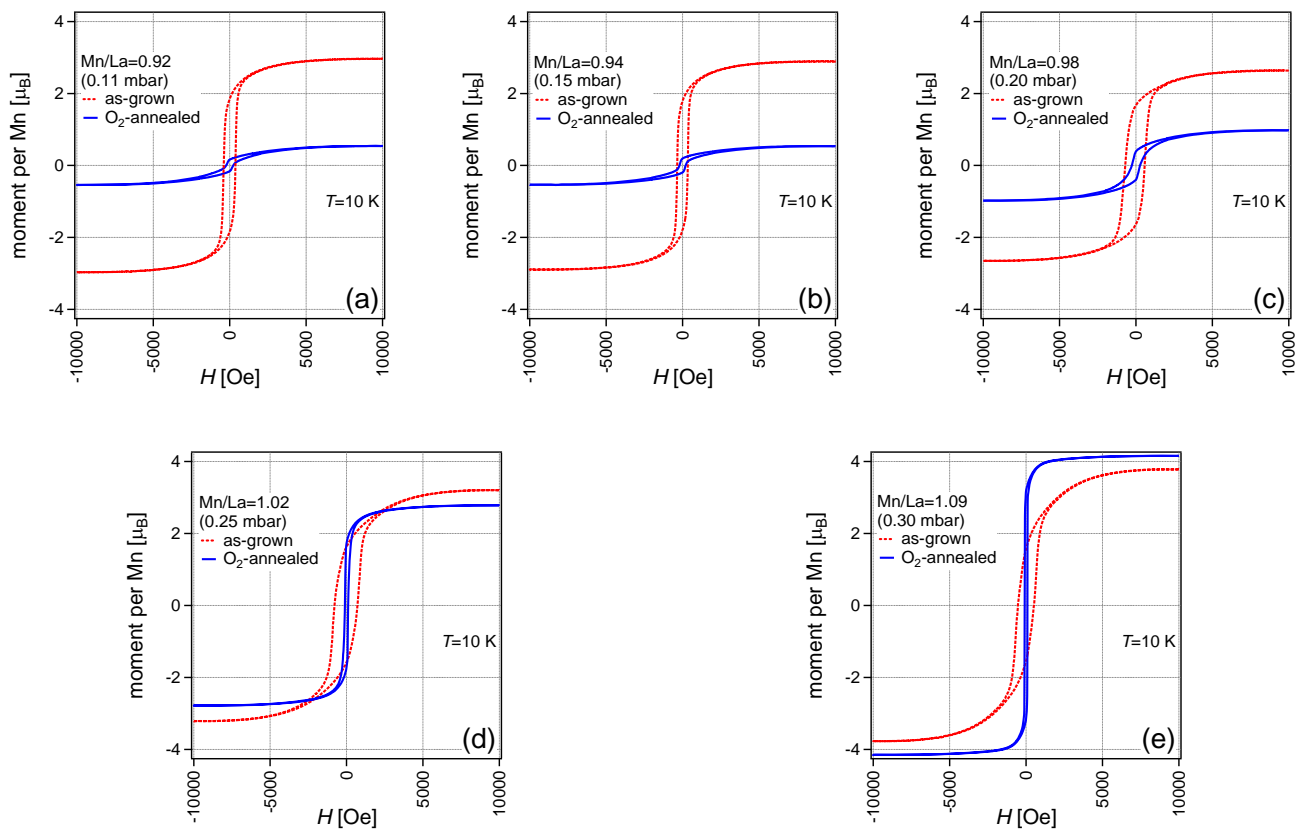

Figure 9 (Color online). Saturation magnetization versus the Mn/La ratio of the LMO films as obtained from the 10 K magnetization loops in Fig. 8.

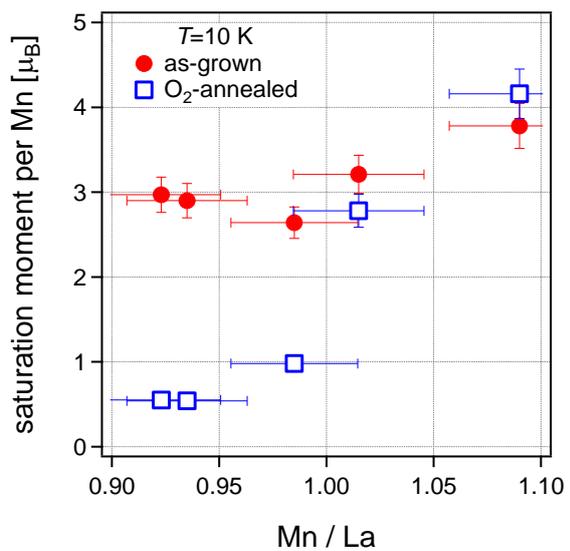

Figure 10 (Color online). Temperature dependence of the real part of the optical conductivity, $\sigma_1$, for **(a)**, **(b)** the most La-deficient and **(c)**, **(d)** the most Mn-deficient LMO films.

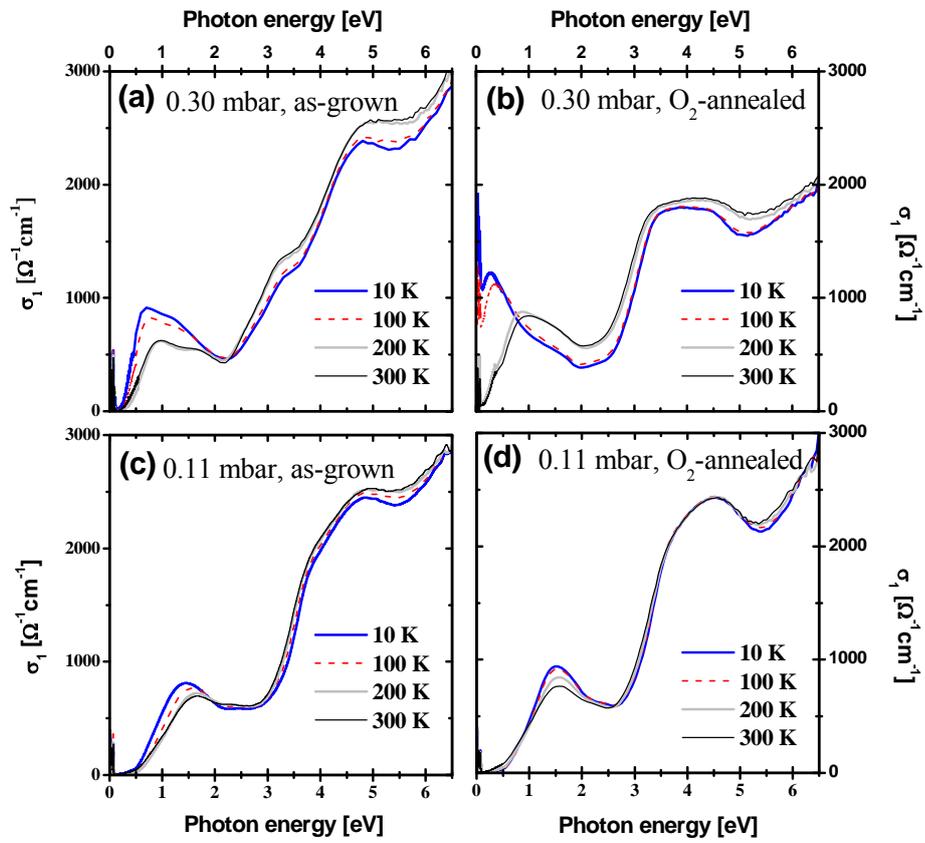

Figure 11 (Color online). **(a)** Comparison of the room temperature optical conductivity spectra of the strongly Mn- and La-deficient LMO films before and after the oxygen annealing treatment. **(b)** Difference spectra of the conductivity in the magnetic state at $T=10$ K and the paramagnetic state at $T=200$ K. Positive (negative) values correspond to a spectral weight gain (decrease). **(c)** Comparison of the room temperature spectra of the as-grown Mn- and La-deficient samples grown at 0.11 and 0.30 mbar of oxygen pressure, respectively. Arrows mark the position of the high-spin and low-spin intersite *d-d* transitions that are strongly affected by the cation vacancies. **(d)** Corresponding spectra for the oxygen-annealed samples.

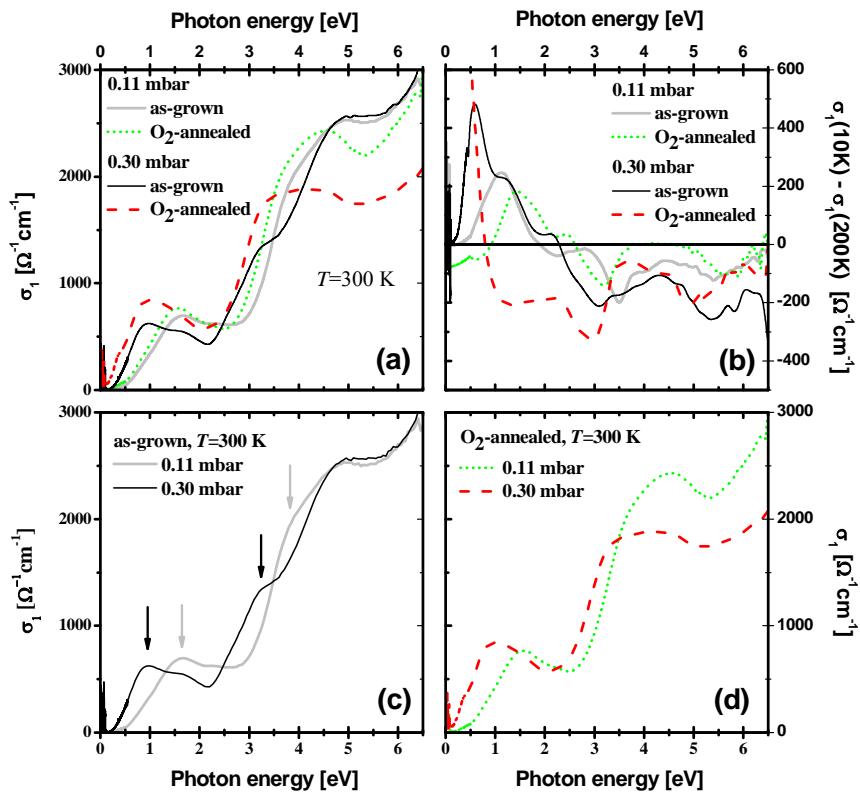